\documentstyle[aaspptwo]{article}

\begin{document}

\title{Discovery of Hidden Blazars}

\author{Feng Ma\footnote{Prc-Mrc 2nd Floor/R9950, University of 
Texas at Austin, 
Austin, TX 78712; feng@astro.as.utexas.edu}, 
Beverley J. Wills\footnote{McDonald Observatory and 
Astronomy Department, University of Texas at Austin,  
Austin, TX 78712}}
\affil{}

{\bf A blazar is believed to exist in every radio-loud
quasar. This is expected in a unified scheme where the differences in 
both optical
and radio observations of radio-loud quasars are the result of
different viewing angles. We have predicted that 
blazars may be detected using emission line ratio variations
caused by variable illumination of gas clouds in the broad
emission line region. In a spectroscopic search of 62 quasars
at a redshift of about 2, we have discovered large ($>$20\%)
variations of the emission line ratios, CIV/CIII] or CIV/Ly$\alpha$, 
when compared with historical data taken 
over 10 years ago. This result is consistent with our prediction, 
and thus supports the unification scheme for radio-loud quasars. }

Quasars are the most luminous of active galactic nuclei (AGN).
About 10\% of all 
cataloged quasars are more luminous at radio than at optical
wavelengths, and thus are classified as radio-loud quasars. 
Their strong radio emission arises
from two jets shooting away from the center 
in opposite directions, terminating in extended radio lobes.
It is believed 
that massive black holes and accretion disks are the central engines of
all quasars ({\it{1,2}}). However, it is unclear what gives rise to 
the powerful radio jets.
It may be that these jets are produced in the vicinity of only the most
massive black holes ({\it{3}}), or perhaps in those with greatest angular momentum
({\it{4}}).
Radio-loud quasars are not all alike. However, the diversity in their radio
structures and optical spectra can be explained by
a unified scheme based on the viewing angle ({\it{5,6}}). 
If we look directly into the jet, we see
strongly  beamed synchrotron radiation with the energy distribution
peaking in the infrared wavelength. 
This synchrotron radiation is highly variable and often dominates 
the light from the quasar's accretion disk and 
broad emission lines. These objects are classified as blazars. 
If the line of sight is a few degrees away
from the jet direction, we can still see a bright radio core, which
dominates the radio power, and this quasar is classified
as core-dominant.  As the viewing angle is increased beyond a few degrees to the jet 
direction, the beamed synchrotron emission in radio through optical wavelengths dims.
When the jet-lobe structure is viewed at a larger angle, 
the infrared through optical emission is  dominated by thermal accretion
continuum and broad emission lines, and
the radio core becomes less bright than the radio lobes. These are the
lobe-dominant quasars. As the viewing angle is further increased, 
the quasar continuum and broad line emission 
are hidden behind an almost edge-on thick dusty nuclear torus or dust in
the host galaxy. The only clue to a hidden quasar may be the extended jets
and radio lobes.  Observationally, this is a radio galaxy.

If the unified scheme ({\it{5,6}}) is correct, every radio-loud quasar 
should harbor a blazar, and we may be able to observe 
signatures of the hidden blazars caused by interactions of the beam
with their otherwise normal host quasars. 
Our models have shown that
the large flux of beamed infrared radiation from a hidden blazar
can heat the broad emission line region (BELR) gas,
enhancing collisionally-excited emission lines such as CIV (154.9 nm)
and SiIV (139.7 nm) ({\it{7}}). A key signature
of this variable heating would be  large variations of the line intensity
ratios, CIV/Ly$\alpha$ or CIV/CIII], 
because Ly$\alpha$, CIII] (190.9 nm) and all Balmer 
lines are much less
affected by the heating. The latter lines are more sensitive 
to hydrogen-ionizing ultraviolet photons that are scarce in the beamed 
steep-spectrum synchrotron radiation.

A large blazar outburst may occur 
once every 10 to 20 years. For example,  3C 279  
showed one outstanding outburst in each of two 20-year monitoring 
periods ({\it{8}}), and each of the outbursts lasted for about one year.  
During the outbursts this blazar was more than 15 times
brighter than usual. Smaller amplitude (a factor of two) variations 
are also common when it is not in outburst. 
If every radio-loud quasar harbors a 
3C 279 type blazar, we expect to see in every radio-loud quasar 
variations of emission
line equivalent width (EW) ratio (CIV/CIII] or CIV/Ly$\alpha$) 
larger than 20\% once
every $\sim$60 years in the observer's frame (for redshift $z{\sim}2$ in 
order for the ultraviolet emission lines to redshift into the optical
wavelength range, and there is a $1{+}z$ time dilation factor). 
Instead of monitoring a few quasars for so many years to look for
the predicted line ratio variations, 
we use an alternative approach, that is, comparing 
new spectra from a sample of 62 radio-loud quasars 
with historical spectra taken over 10 years ago.
The quasars in our
sample have a redshift $1{<}z{<}3$ and absolute magnitudes 
${-}31{<}M_{\rm B}{<}{-}25$. 
Since 3C 279 spent $\sim$5\% of its lifetime in outburst, 
we would expect to catch three outbursts in our sample
during each of the two epochs, or six events with EW CIV/CIII] or 
EW CIV/Ly$\alpha$ variations larger than 20\%. 
Because of the large range in blazar luminosity and variability, 
the event rate may be higher or lower than estimated here.

Earlier studies of AGN variability have focused on
nearby, low-luminosity, radio-weak  objects 
with known continuum variability, such 
as NGC 5548, for which it is found ({\it{9}}) that Ly$\alpha$, CIV 
and CIII] lines vary with similar amplitudes
in response to variations of the normal, 
non-synchrotron continuum (that may arise in an accretion disk, for example). 
The amplitude of line variations is
usually less than half that of the continuum; this is also true for 
Balmer lines in a sample of 
low-redshift ($z{\sim}0.1$) quasars ({\it{10,11}}), and
a sample of 
$z{=}0.14{-}0.59$ radio-loud quasars, including both core-dominant 
and lobe-dominant objects ({\it{12}}). 
Systematic spectroscopic monitoring of
luminous high-redshift quasars has not been done before.
From photometric studies of large samples
of quasars it is found that those of higher luminosity are 
less variable ({\it{13--15}}). 
For a quasar in the luminosity range 
of our sample (${-}31{<}M_{\rm B}{<}{-}25$), 
when measured at two epochs, the typical variation
is less than 0.2 magnitude
({\it{13,14}}). 
The expected emission line variations are ${<}10$\% in response
to a 0.2 magnitude normal continuum variation. The line ratio
variations are even smaller  ({\it{9,16}}).

Most of the new spectra were obtained over the years 1998 to 2000,
using the Large Cassegrain 
Spectrograph on the 2.7-m Harlan J. 
Smith telescope at McDonald Observatory.  
We typically used a 2$^{\prime\prime}$ slit for 
long exposures (e.g. 5400 seconds for objects of 18.5 visual magnitude),
followed by a short (600 second) exposure with a wide slit 
(8$^{\prime\prime}$) to correct the narrow-slit spectra. 
The  seeing was typically 2--3$^{\prime\prime}$ and thus
an 8$^{\prime\prime}$ slit exposure is sufficient
to correct the shape of narrow-slit
spectra for atmospheric refraction losses. 
During each night usually five standard stars
were observed using the 8$^{\prime\prime}$ slit to 
calibrate quasar continuum shapes.
Each of the spectra is wavelength calibrated using argon and neon
lamps. Many spectra were
taken under non-photometric conditions and thus 
are not on an absolute flux density scale.

Most historical spectra we used, such as those from ({\it{17}}),  
were obtained using narrow slits and are not absolutely flux calibrated. 
Hence, we focus on comparing emission line EW ratio variations. 
Therefore we used direct division of two spectra to reveal the variability
of emission lines relative to the local continuum [e.g. ({\it{18}})]. 
A featureless division spectrum indicates that there are probably 
no variations in the continuum or emission lines, because terrestrial 
clouds only cause a gray suppression of the spectra, and 
wavelength-dependent
slit losses resulting from atmospheric dispersion cause a 
low order continuum shape difference. 
We divide spectrum 1 at one epoch, by the spectrum 2 at another epoch.
Avoiding regions of emission lines, we fit a low order curve to this
division spectrum, and multiply spectrum 2 by this curve.  Thus we
normalize the continua, removing the effects of wavelength-dependent
continuum variation.
Information about changes in the ratios of emission line strengths
is preserved unless there is a real variation in the continuum shape.
Very little is known on how the continuum shape varies in a quasar
spectrum. However, some clues can be inferred from NGC 5548, 
which appears to be $\sim$17\% more variable at 135.0 nm 
than at 184.0 nm in the rest frame while the average continuum variation 
amplitude is a factor of two ({\it{15}}). This
would cause a 17\% apparent emission line EW ratio (Ly$\alpha$/CIII])
variation in our method of analysis. We note that quasars
in our sample are much less variable than NGC 5548, and hence
the continuum shape variation is likely to be less than 17\%.

For half of the objects in our sample the spectra
at the two epochs match well (Table 1, class C). This is consistent
with photometric findings that quasars at this
redshift and luminosity range are generally not highly variable, 
so our comparison method appears tenable.
Most of the class D 
objects, which are core-dominant and known to be
more variable than lobe-dominant objects ({\it{18}}), 
could cause the apparent proportional 
line variations using our method of comparison. 
Large line variations are  seen only in CIV lines (Fig. 1), 
consistent with our predicted behavior for 
quasars with hidden blazars. 

MRC 0238$+$100 [$z{=}1.83$ ({\it{17}}), 
$M_{\rm B}{=}{-}27.7$ ({\it{19}})] has shown an 
increase of 70\% in its CIV line strength between
2 December 1986 and 10 November 1999. 
A SiIV (139.7 nm) line, not present in the historical spectrum, 
appeared in our new spectrum. 
The SiIV line is more sensitive to infrared
heating than the CIV line, and the intensity of the SiIV line 
may increase by 100\% according to our
models ({\it{7}}). The appearance of the SiIV line in our
new spectrum suggests that the CIV enhancement is due to infrared heating from
the additional blazar continuum. 
The low signal-to-noise ratio near the CIII] line means that 
we can only constrain the CIII] variation to be $<$30\%. 
Hence, the variation in emission line EW ratio in this case is
$>$30\% for CIV/CIII] over the time interval of 13 years. 
From the digitized (red plate) Palomar observatory sky surveys I and II 
({\it{20}}) we find for MRC 0238$+$100 
a differential variation of $0.01{\pm}0.20$ magnitude between
epochs 1954 
and 1990. Comparing our four new observations on 
22 November 1998, 10 November 1999, 2 March 2000, and 1 January 2001,
a time interval of 2.1 years (or 0.74 years in this quasar's rest frame), 
reveals no significant continuum or emission line variations. 
We suggest that the hidden blazar inside MRC 0238$+$100
is currently in an outburst, maintaining strong CIV and SiIV line emission
with little change in the observed continuum
or CIII] emission line. 
The low activity state of the hidden blazar in 1986 is 
supported by an earlier observation in 1976 ({\it{21}}), 
when MRC 0238$+$100 had line-to-continuum ratios (defined as flux ratios
from the line peak to the continuum level. Unfortunately the spectrum
was not published) of 2.50 for CIV, and
1.53 for CIII]. These observed line strengths are  
consistent with the 1986 values 
while in our new spectrum the line-to-continuum ratios are 3.8 for CIV and 
1.6 for CIII], as can be measured from Fig. 1.

UM 556 [$z{=}2.39$, $M_{\rm B}{=}{-}28.4$ ({\it{19}})]
has a 70\% stronger CIV emission line in the spectrum 
of 5 June 2000 compared with the spectrum of 6 March 1988. 
The Ly$\alpha$ line is $\sim$10\% 
stronger in 2000 than in 1988. The SiIV line also appears to be
stronger in 2000 but with less certainty due to its being 
a weak line in these spectra.  
The line EW ratio variation is thus 55\% for CIV/Ly$\alpha$. 

If we assume that the blazar outbursts do not last longer than 
one year in the rest frame as suggested by observations of nearby blazars, 
we can predict that the CIV 
line in MRC 0238$+$100 and UM 556 will drop significantly within the 
next three years.  
Even though the hidden blazar's continuum may vary on times scales of
less than several months or even days, the observed emission lines are
unlikely to vary on these time scales, because the light travel time 
across the BELR [several light months in high-luminosity quasars
({\it{11}})] will result in smearing out such variations.

PKS 0038$-$019 [$z{=}1.67$ ({\it{17}}),
$M_{\rm B}{=}{-}26.9$ ({\it{19}})] shows a
CIV line  43\% stronger in 1986 than in 1999, 
while  its CIII] line is 18\% stronger in 1986 than in 1999. 
The line EW ratio (CIV/CIII]) in 1986 is thus 21\% larger than in 1999.   
The emission line intensity variation of CIV 
is seen in the red wing, suggesting that  
the part of BELR gas in the beam is moving
away from us. 
This shows the potential of the technique to probe the polar
regions of the quasar BELR. However, this case is not as convincing  as in 
MRC 0238$+$100 because there is
uncertainty in matching the continuum at the blue end of
the spectra and the historical spectrum 
did not include the SiIV line.  
We also derive from the digitized Palomar observatory sky surveys ({\it{20}}) 
a differential photometric variation of $0.11{\pm}0.45$mag. The large error bar
may be due to the plate quality and the slight
difference in the bandpasses at the two epochs.

For PKS 0424$-$13 [$z{=}2.17$, 
$M_{\rm B}{=}{-}28.6$ ({\it{19}})], the spectrum obtained
on 16 February 1990 shows the CIV line to be  25\% stronger 
than on 20 December 1998, while Ly$\alpha$ is $\sim$10\%
stronger in 1990 than in 1998. The line EW ratio CIV/Ly$\alpha$ 
is thus 14\% larger in 1990 than in 1998. In addition, the CIII] line 
shows little variation in this object. The CIV line variation
is thus constrained from both ends of the spectrum, ruling
out the possibility that the observed line variation is caused
by continuum shape variation in this object.

Both MRC 0238$+$100 and PKS 0038$-$019 are lobe-dominant with clear 
double-lobed radio structures ({\it{22}}).  
The ratios of core to total radio flux density
 of  MRC 0238$+$100 and PKS 0038$-$019  at rest frame 5 GHz
are 0.098 and 0.067, respectively ({\it{23}}). 
The lobe-dominant radio structure of PKS 0424$-$13 
is implied by its steep radio spectrum ({\it{24}}). 
We do not expect to see the variable 
blazar continuum at such large implied viewing angles
away from the beam. While some large apparent
emission line variations have been reported 
in core-dominant quasars with large continuum
variations [e.g., ({\it{18}}) reported 
an EW change of 68\% in CIV and 82\% in CIII] for 3C446], no large line ratio 
variations comparable to the cases of MRC 0238$+$100 and UM 556 have been
reported so far,  except for the low luminosity AGN NGC 5548
[e.g., ({\it{25}}), where a HeII (468.6 nm) flare was 
interpreted as an accretion event]. We also note that
because 90\% of quasars are radio quiet, some of them have
inevitably been observed more than once, yet no large
line ratio variations have been published ({\it{26}}). 

Our observations offer support for
the unified scheme for radio-loud quasars. The emission line variations
provide the most direct evidence for the existence 
of a  violent blazar in every radio-loud quasar. Similar ideas
can be applied to other jet-disk systems such as 
gamma-ray bursts and proto-stars. In addition, 
the disk-wind model for the broad line emitting clouds
with the winds blowing off the accretion disk by radiation 
pressure  ({\it{29}})
is challenged because it does not include any
gas in the polar regions. Another widely adopted model, 
the stellar atmosphere model
for the BELR ({\it{30}}) has been challenged by emission
line profile studies ({\it{31}}). 
Hence, other explanations for the origin
of BELR gas may be needed. 
Finally, the results from this work suggest that 
radio-loud quasars should be excluded when studying
cosmology via the Baldwin effect [the inverse relation between EW(CIV) and
continuum luminosity,  ({\it{32}})]. 
Reducing the scatter in this relationship
is critical
for using quasars as luminosity indicators ({\it{33}}), 
while radio-loud quasars introduce extra scatter 
as a result of their hidden blazar activity.

\clearpage

\begin{table}
\begin{center}
\caption{Classification of emission line EW ratio variability in the sample. 
The  five  classes are indicated
in column 1. Column 2 gives the number of objects in each
class. The number of core-dominant objects is given in
the parentheses.}
\begin{tabular}{lccc}
\\
\hline \hline
\\
Class & \#(c) & ~~~~~~~~Description~~~~~~~~ & Interpretation \\
\\
\hline
\\
A  &  3(1)  & large relative CIV   & outbursting hidden  \\
   &     &  variations  ($>$20\%)   & blazars revealed   \\
\\
B  &  9(3)  & smaller relative CIV  &  hidden blazars active\\
   &     &  variations  (10--15\%)   &   \\
\\
C  &  33(7) & no $>$10\% variations & high-$z$ and high-$L$  \\
   &     & in any lines       & quasars not very variable \\
\\
D  &  15(10) & all lines vary & continuum variation \\
   &     & in proportion     & at different epochs; \\
   &     &                   & emission line response to\\
   &     &                   &  normal quasar continuum\\
\\
E  &  2(1)  & 10--15\% variations & continuum more\\
   &     &   in Ly$\alpha$  & variable in the blue \\
   &     &  but not CIV              & than in the red\\
\\
\hline
\end{tabular}
\end{center}
\end{table}

\clearpage

\begin{figure}
\caption{Comparison of spectra at two 
epochs. The continua at different epochs have
been scaled to the same
level. Red solid lines are new spectra taken during 1998--2000. Blue dotted
lines are historical spectra taken during 1986--1988. 
Subtraction spectra (green lines) are plotted in the same panels. 
The flux level is plotted in $\mu$Jy [1 jansky (Jy)$=10^{-26}$ W m$^{-2}$ Hz$^{-1}$]. 
 In the lower
panels the division 
spectra (before scaling) and the fitting curves are plotted together in 
arbitrary units. 
MRC 0238$+$100, UM 556, PKS 0038$-$019 and PKS 0424$-$13 show 
significant CIV variations with little 
CIII] or Ly$\alpha$ variation.
PKS 2354$+$14 and 4C 05.34 are examples of class C objects whose
spectra taken at a time interval of over 10 years using different
telescopes and instruments are in excellent agreement.}
\end{figure} 


\clearpage

\begin{figure}[hp]
\plotone{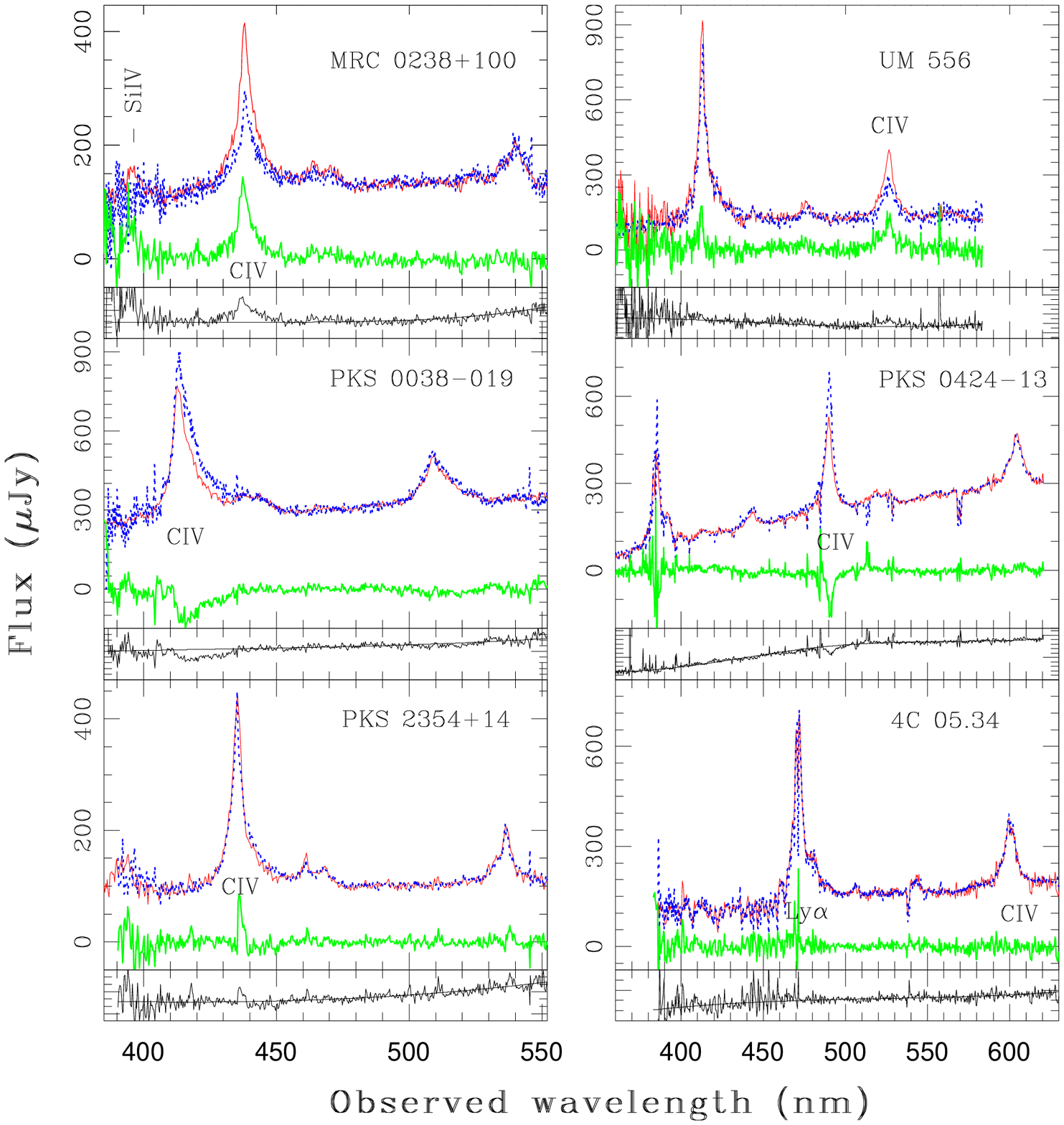}
\end{figure}

\end{document}